\documentclass{article}
\usepackage{latexsym}
\usepackage[psamsfonts]{amssymb}
\usepackage{amsmath}
\newcommand{\ie}{{\em i.e$.$} }

\newcommand{\alp}{\alpha}

\newcommand{\del}{\delta}

\newcommand{\iot}{\iota}

\newcommand{\sig}{\sigma}

\title{{\Large {\bf 
Noether Theorems and Reality of Motion}}}

\author{Marcella Palese  and Ekkehart Winterroth \\
{\footnotesize Department of Mathematics, University of Torino} \\ {\footnotesize via C. Alberto 10, I-10123 Torino, Italy } \\ 
 {\footnotesize {\sc e-mail: marcella.palese@unito.it}}}

\date{}

\begin{document}
\maketitle

\begin{abstract}

We will read, through the Emmy Noether paper and the two concepts of  `proper' and `improper' conservation laws, the problem, posed by Hilbert, of the nature of the law of conservation of energy in the theory of General Relativity. Epistemological issues involved with the two kind of conservation laws will be enucleate.



\end{abstract}

\section{Introduction}


In a recent work\cite{Kos11}, Y. Kosmann-Schwarzbach, besides providing a fine French and English translation of the celebrated 1918 Emmy Noether paper \cite{Noe18},  thoroughly analyzed the inception and the influence in Physics as well as the historical developments of Noether Theorems during the XXth Century. 
In that work the author stresses the surprisingly small number of references to Noether before 1950 and in  particular the astonishing absence of citations dealing with invariance problems in standard treatises, now classical textbooks,
on the variational calculus.
Particularly interesting, there it is noticed how contemporary physicists such as Pauli and even Weyl somehow seemed to overlook the significance of her work, quoting instead the subsequent work of Bessel-Hagen\cite{BeHa21} on conservation laws in electrodynamics. Sure, Noether's paper implied  epistemological issues difficult to stand with by the community of physicists for which equations, or dynamics, was a sort of `totem', since from Modern Age on Physics {\em \bf  was} dynamics.
 Emmy Noether is interested in conservation laws, she is  interested in what changes insomuch as this is an outcome of  what remains unchanged \cite{FrPaWisisfa12};
 which can perhaps explain why Noether contemporary and even some today's physicists do not feel `comfortable' especially with Noether's Second Theorem and the related concept of an `off shell'' conserved quantity. 

Philosophical considerations about space, time and motion are strongly based on what Physics should be, and that brings us back to  natural  philosophy. 
As well known, in his poem about the Nature\cite{Par06}  Parmenides points out the question about `Be' and `Not' and about the illusory character of past and future, \ie of time (\cite{Par06} Fragm. $8$).
He poses the necessity  of a Being full and steady and how things, such as, in particular, {\em {\bf changing of place}} are just {\bf labels} fixed by human beings (\cite{Par06} Fragm. $19$);
however, things which appear needed to really be, being totally in each sense (\cite{Par06} Fragm.  $1$):
Here  the  question of a justification of a Becoming which is Being for a metaphysical reason is posed and therefore of a possible metaphysical distinction among an ``illusory'' and a ``real'' physical world; in particular, the ``real'' is seen as the {\bf duality} of ``light'' and ``darkness'' because with no one of them is the `Not' (\cite{Par06} Fragm. $9$).

Aristoteles natural philosophy dominates the scene up to the
birth of experimental sciences with Galilei's {\em Dialogo sopra i due massimi sistemi del mondo} (1632) and {\em Discorsi e dimostrazioni matematiche intorno a due nuove scienze attinenti la mecanica e i moti locali} (1638), whereby the baroque `wonder'  (epitomized by the Galilei's experiment of the brachystochrone) discovers the concept
of law of motion: Descartes, Newton, Leibniz provide the analytical geometry and the
development of calculus necessary to the development of differential geometry by means of which the
{\em motion is given as parametrized trajectories (change of
space coordinates with time)}.

\section{The Calculus of Variations}

The problem of  reality of dynamics will be posed as the need to justify  why dynamics is given in a certain way:  modern Mathematical Physics as initiated by Fermat and developed by Euler
would have been conceived as a theory of what changes (equations of motion) sorted out by what remains unchanged (the action).

\subsection{The problem of minimizing time and motion}

Fermat postulates that Nature should follow the path requiring {\bf the shorter time}.
In his `Methodus ad disquirendam maximam et minimam' (\cite{Fe1891}),  for the study of refraction of a light ray in an optical medium, in particular in  `Synthesis  ad Refractiones'  Fermat writes (boldface and emphasis by the authors):
\begin{quote}
{\footnotesize

 ``Demonstratio  nostra unico nititur postulato: {\em naturam operari per modos et vias faciliores et expeditiores}. Ita enim $\alp {\bar{\iot }}\tau\eta\mu\alp$ concipiendum censemus, non, ut plerique, {\em naturam per linea brevissimas semper operari}.
 
 {\bf Ut enim Galilaeus}, dum motum naturalem gravium speculatur, {\bf rationem ipsius non tam spatio quam tempore metitur}, pari ratione {\bf non brevissima spatia aut lineas, sed quae expeditius, commodius et breviori tempore percurri possint, consideramus.}"
 }
 \end{quote}
In \cite{Eu1744} Euler generalizes the metaphysical insight of minimizing time by stating a  {\bf principle of minumum motion} (boldface and emphasis by the authors):
\begin{quote}
{\footnotesize 
{\em Spectari autem possimun debet effectus a viribus sollicitantibus oriundus; qui cum in motu corporis genito consistant, {\bf veritari consentaneum videtur hunc ipsum motum}, seu potius aggregatum omnium motuum qui in corpore projecto insunt, {\bf minimum esse debere}.} 
}
 \end{quote}
 \subsection{Lagrange's concept of  `virtual displacement'}
In his M\'echanique Analitique 
(1788), 
Lagrange denies a metaphysical meaning of the principle of minimum action.
He creates his  Analytical Mechanics as a formal calculus by introducing the concept of  `virtual displacement':
\begin{quote}
{\footnotesize
En g\'en\'eral, il faut remarquer relativement aux {\em variations}, qu'elles ne se rapportent qu'\`a l'espace \& non \`a la dur\'ee, ensorte que dans les diff\'erentiations marqu\'ees par $\del$ la variable $t$, qui  repr\'esente le temps devra toujours \^etre regard\'ee comme constante.
} 
\end{quote}
Therefore, Calculus of Variations becomes just
 an {\em analytical  technique} to obtain equations of motions in Mechanics from the principle of conservation of energy.

\section{Noether $1918$ paper and the problem of energy in field theory}

Since then,  a long term ``revolution''  in the Calculus of Variations, due, among the others, to Hamilton, Jacobi, Hilbert, Noether and Lepage, which poses the roots in Maxwell theory of electromagnetic field, breaks out with the problem of conserved quantities in field theory.
If fields depend equally from space and time, what exactly should be energy as a product of invariance in space-time?

In particular, concerning the general theory of gravitation, although the gravitational field equations were global, the associated conservation laws found by Einstein 
were not (think of the well known energy-momentum {\em pseudo-tensor}). 

In the introduction of {\em Invariante Variationsprobleme} (1918), Noether wrote\footnote{`For those differential equations that arise from variational problems, the statements that can be formulated are much more precise than for the arbitrary differential equations that are invariant under a group, which are the subject of Lie's researches'. Translated from German by Y. Kosmann-Schwarzbach and B.E. Schwarzbach \cite{Kos11}.}  (boldface   by the authors):
\begin{quote}
{\footnotesize 
\"Uber diese {\bf aus Variationsproblemen entspringenden} Differentialgleichungen lassen sich viel pr\"azisere Aussagen machen als \"uber beliebige, eine Gruppe gestattende Differentialgleichungen, die den Gegenstand der Lieschen Untersuchungen bilden.
}
\end{quote}
The relevance of the study of differential equations generated by an invariant  variational problem in its whole is in the issue of a major refinement in the results: to symmetries of equations could correspond conservation laws which have a non variational meaning and thus could not be characterized in a similar  precise manner\cite{FrPaWisisfa12}.

While physicists prefer to study symmetries of equations, because they are transformations of the space leaving invariant the description of field equations which {\em describe the changing} of the field in base space, the formulation of an invariant  variational principle (\ie a principle of {\em stationary} action) keeps account of {\em both} what ({\em and how}) changes and what ({\em and how}) is conserved. Mathematically, Euler-Lagrange field equations are `adjoint' to stationary principles up to conservation laws: 
a contemporary mathematical formulation of the duality between {\em Being} and {\em Becoming} 
(light and darkness in Parmenides' words).

From an epistemological point of view Emmy Noether considers field equations insomuch as they are generated by equating to zero what she calls the Lagrange expressions. She mentions `equations' only few times in her paper. In particular, she underlines that Euler-Lagrange equations derive by the variational principle by requiring the boundary term vanishes (they are given by equating to zero `Lagrange expressions').
She, however, is very much more interested in the boundary term than in the equations; in fact she is investigating conservation laws.

Her epistemological point of view is: we are interested in what remains unchanged (conserved quantities). There exists equations, but we are not interested directly to them, rather we are interested to how Lagrange expressions are important insomuch as they can say something about what remains unchanged. In particular she concentrates on the `work' term obtained by contraction of the Lagrange expressions with the variation vector field. Therefore, all her work is based on the assumption that we are not along solutions of Euler-Lagrange equations (the work term would be zero). Therefore she {\em wants} to be, how physicists say,  {\em `off shell'}. 
\subsection{Noether's concept of variation field}

The second main epistemological novelty she introduces is what `virtual displacements' (\ie variations) should be in field theory. She takes as variations vertical parts of infinitesimal generator of invariant transformations of the Lagrangian. In fact the first passage is to split the work term in a summand containing vertical parts of generators of the invariance transformation contracted with Lagrange expressions  and a summand going under a divergence which contains the horizontal part of generators of the invariance transformation contracted with the Lagrangian. This piece going under a divergence is the contribution to the Noether current which does not come from the momentum. The latter is obtained by applying the standard variation calculus {\em with variations which are generated by the invariance transformation}.
This gives a `work term' plus a further divergence term (the momentum term) which sum up with the contribution due to the horizontal part of the symmetry.  

\section{The two statements and their epistemological implications}

In what follows we shall examine the two Theorems she states (for convenience of the reader as translated in English in \cite{Kos11}):
\begin{quote}
{\em I.} If the integral $I$ is invariant under a [group] $G_\rho$, then there are $r$ linearly independent combinations among the Lagrangian expressions which become divergences -- and conversely, that implies the invariance of I under a [group] $G_\rho$. The theorem remains valid in the limiting case of an infinite number of parameters.
\end{quote}
The meaning of this first statement is in relating, by invariance properties of the Lagrangian, Lagrangian expressions (thus `equations') with conservation laws; that is: {\em expressing what changes by what remains unchanged}. 
The First Theorem provides us with the following  epistemological assertion: equations obtained from invariant Lagrangians can be related to conserved quantities, in particular {\em  what changes can be related with what remains unchanged if and only if it derives variationally from an invariant Lagrangian}. 
\begin{quote}
{\em II.} If the integral $I$ is invariant under a [group] $G_{\infty \rho}$ depending on arbitrary functions and their derivatives up to order $\sig$, then there are $\rho$ identities among the Lagrangian expressions and their derivatives up to order $\sig$. Here as well the converse is valid.
\end{quote}
She also adds:
\begin{quote}
For mixed groups, the statements of these theorems remain valid; thus one obtains identities \footnote{The authors of the present note would prefer the word `dependencies' as translation for {\em Abh\"angigkeit}} as well as divergence relations independent of them.
\end{quote}
The Second Theorem is a further investigation of the `work term' containing the Lagrange expressions. Once we have realized that such a term can be related with conserved quantities, under which conditions can we further transform it in such a way that it becomes a divergence itself ({\em independently from the invariance of the Lagrangian})? The answer she finds is: yes it is always possible to make the `work term' become  a divergence if and only if the group of symmetry transformations is an infinite continuous group (\ie depends on a given number of functions rather than just parameters). In this case the existence of identities among Lagrange expressions and their derivatives guarantees that the `work term'  reduces to a further divergence.

Furthermore, as a consequence of invariance we can deduce that  a divergence is identically zero (also off shell) providing what she calls an `improper' conservation law.

The Second Theorem is very strong: taking as variations the generators of transformations of independent and dependent coordinates of that particular type, always guarantees that contractions of Lagrangian expressions with vertical parts of transformations become divergences and {\em vice versa}. If, and only if, the Lagrangian is invariant under such transformations therefore we can further characterize such divergences as conserved quantities also off shell (so-called `strong' conserved quantities). 
Notice that this means that when such symmetries are given, due to Noether identities we can always transform the dynamical content in a component of the strong conserved quantity. 

The Second Theorem therefore attributes a strong meaning to such kind of symmetries and therefore provides us with  the epistemological assertion: equations obtained from invariant Lagrangians with respect to such kind of symmetries {\em always} can be put in the form of an `improper' conservation law, more specifically 
{\em  what changes {\bf ``is''} part of what remains unchanged if and only if it derives variationally from an invariant Lagrangian with respect to an infinite group of transformation as above}. 

Therefore dynamics is `real' if and only if it derives from an invariant Lagrangian with respect to such kind of symmetry transformations.

Theorem II. was inspired by a question of Hilbert on the nature of the law of conservation of energy in General Relativity.
Since general transformations of coordinates in General Relativity form a group as $G_{\infty \rho}$; in a dedicated Section she shows how this conservation law, because of Theorem II.,  is in fact a divergence vanishing identically, and therefore an `improper' conservation law.
Physical theories of the fundamental interactions have been all stated as gauge-natural field theories in terms of invariant Lagrangians with respect to symmetries as stated in the Second Noether Theorem. The epistemological content of the Noether Theorems became therefore a basic requirement for meaningfulness of a physical theory.
Notice also that a vertical variation (\ie a virtual displacement) is not yet enough to perform variations, meaningful variations instead are the ones obtained by invariance group of transformations of the Lagrangian.

\section*{Acknowledgments}
Research supported by Department of Mathematics University of Torino through local research project {\em Metodi Geometrici in Fisica Matematica e Applicazioni 2013-2015}.

\end{document}